\documentstyle[amssymb,aps,preprint]{revtex}

\begin{document}
\title{First principles theory of fluctuations in vortex liquids and solids.}
\author{Baruch Rosenstein}
\address{National Center for Theoretical Studies and Electrophysics Department,\\
National Chiao Tung University, Hsinchu, Taiwan, R.O.C.}
\maketitle
\date{October 10, 1998 }

\begin{abstract}
Consistent perturbation theory for thermodynamical quantities in type II
superconductors in magnetic field at low temperatures is developed. It is
complementary to the existing expansion valid at high temperatures.
Magnetization and specific heat are calculated to two loop order and compare
well to existing Monte Carlo simulations and experiments.
\end{abstract}


Thermal fluctuations play much larger role in high $T_{c}$ superconductors
then in the low temperature ones because the Ginzburg parameter $Gi$
characterizing fluctuations is much larger. In the presence of magnetic
field the importance of fluctuations in high $T_{c}$ superconductors is
further enhanced. Strong magnetic field effectively suppresses long
wavelength fluctuations in direction perpendicular to the field reducing
dimensionality of the fluctuations by two \cite{Nelson}. Under these
circumstances fluctuations influence various physical properties and even
lead to new observable qualitative phenomena like vortex lattice melting
into vortex liquid far below the mean field phase transition line. It is
quite straightforward to systematically account for the fluctuations effect
on magnetization, specific heat or conductivity perturbatively above the
mean field transition line using Ginzburg - Landau description \cite{Tinkham}%
. However it proved to be extremely difficult to develop a quantitative
theory in the interesting region below this line, even neglecting
fluctuation of magnetic field and within the lowest Landau level (LLL)
approximation.

To approach the region below the mean field transition line $T<T_{mf}(H)$
Thouless \cite{Thouless} proposed a perturbative approach around homogeneous
(liquid) state was in which all the ''bubble'' diagrams are resummed. The
series provide accurate results at high temperatures, but for LLL
dimensionless temperature $a_{T}\equiv (T-T_{mf}(H))/(TH)^{2/3}\lesssim -2$
become inapplicable (see dotted lines on Fig.2,3 which represent successive
approximants). Generally attempts to extend the theory to lower temperature
by Pade extrapolation were not successful and require additional external
information about the low temperatures \cite{Moore1}. Alternative, more
direct approach to low temperature fluctuations physics might have been to
start from the Abrokosov solution at zero temperature and then take into
account perturbatively deviations from this inhomogeneous solution.
Experimentally it is reasonable since, for example, specific heat at low
temperatures is a smooth function and the fluctuations contribution
experimentally is quite small. This contrasts sharply with theoretical
expectations.

Long time ago Eilenberger calculated spectrum of harmonic excitations of the
triangular vortex lattice \cite{Eilenberger}. Subsequently Maki and Takayama 
\cite{Maki} noted that the gapless mode is softer then the usual Goldstone
mode expected as a result of spontaneous breaking of translational
invariance. The propagator for the ''phase'' excitations behaves as $%
1/(k_{z}^{2}+c(k_{x}^{4}+k_{y}^{4}))$. The influence of this unexpected
additional ''softness'' apparently goes even beyond enhancement of the
contribution of fluctuations at leading order. It leads to disastrous
infrared divergencies at higher orders rendering the perturbation theory
around the vortex state doubtful. For example contributions to energy
depicted on Fig. 1A and 1D are respectively $\log ^{2}(L)$ and $L^{4}$
divergent ($L$ being an IR cutoff) and at higher orders divergencies get
worse. Also qualitatively one argues \cite{Moore2} (in a way similar to that
used frequently to understand the Mermin - Wagner theorem \cite{Mermin})
that lower critical dimensionality for melting of the Abrikosov lattice is $%
D=3$ and consequently vortex lattice in clean materials exists in the
thermodynamic limit only at $\ T=0$. One therefore tends to think that
nonperturbative effects are so important that such a perturbation theory
should be abandoned \cite{Ruggeri}\ and it was abandoned. However a closer
look at the diagrams like Fig.1D (see some details below) reveals that in
fact one encounters actually only logarithmic divergencies. This makes the
divergencies similar to so called ''spurious'' divergencies in the theory of
critical phenomena with broken continuous symmetry. In that case one can
prove \cite{David} that they exactly cancel at each order provided we are
calculating a symmetric quantity.

In this note I show that all the IR divergencies in free energy or other
quantities invariant under translations cancel to the two loop order. I
calculate magnetization and specific heat to this order, interpolate with
existing high temperature expansion and compare with Monte Carlo (MC)
simulation \cite{Sasik} of the same system and experiments. Qualitatively
physics of fluctuating $D=3$ GL model in magnetic field turns out to be
similar to that of spin systems (or scalar fields) in $D=2$ possessing a
continuous symmetry. In particular, although within perturbation theory in
thermodynamic limit the ordered phase (solid) exists only at $T=0$, at low
temperatures liquid differs very little in most aspects from solid. One can
effectively use properly modified perturbation theory to quantitatively
study various properties of the vortex liquid phase.

The GL free energy is 
\begin{equation}
G=\frac{\hbar ^{2}}{2m_{ab}}|(\vec{\nabla}-\frac{ie^{\ast }}{c}\vec{A})\psi
|^{2}+\frac{\hbar ^{2}}{2m_{c}}|\partial _{z}\psi |^{2}+a|\psi |^{2}+\frac{b%
}{2}|\psi |^{4}  \label{energy}
\end{equation}
Here $\vec{A}=(-By,0)$ describes a nonfluctuating almost constant magnetic
field in $c$ direction. Within the LLL approximation $\psi $ can be expanded
in a basis of quasimomentum ${\bf k}$ eigenfuntions 
\begin{equation}
\psi (x)=v\varphi (x)+\frac{1}{2\pi }\int d^{2}k\varphi _{{\bf k}}(x)\sqrt{%
\frac{\gamma _{{\bf k}}}{2\left| \gamma _{{\bf k}}\right| }}\left(
O_{k}+iA_{k}\right)   \label{shift}
\end{equation}
\[
\varphi _{{\bf k}}=\sqrt{\frac{2}{\sqrt{\pi }a_{\bigtriangleup }}}%
\sum\limits_{l=-\infty }^{\infty }\exp \left\{ i\left[ \frac{\pi l(l-1)}{2}+%
\frac{2\pi }{a_{\bigtriangleup }}l(x-k_{y})-xk_{x}\right] -\frac{1}{2}%
(y+k_{x}-\frac{2\pi }{a_{\bigtriangleup }}l)^{2}\right\} 
\]
Unit of length will be the magnetic length $l_{H}\equiv \sqrt{c\hbar \text{/}%
eH}$ and $a_{\bigtriangleup }\equiv \sqrt{4\pi /\sqrt{3}}$ is the lattice
spacing. The ${\bf k}=0$ component $\varphi _{0}(x)\equiv \varphi (x)$ is
''a vacuum'' with its VEV denoted by $v$. The integration is over Brillouin
zone. Instead of one complex field two real fields $O$ and $A$ were
introduced. They are somewhat analogous to acoustic and optical phonons in
usual solids with some peculiarities due to strong magnetic field studied in
detail by Moore \cite{Moore1}. For example the $A$ mode corresponds to shear
of the two dimensional lattice. Substituting eq.(\ref{shift}) into free
energy, quadratic terms in fields define propagators, while cubic and
quartic are interactions. The phase factors containing a function $\gamma _{%
{\bf k}}\equiv \int_{x}\varphi ^{\ast }(x)\varphi ^{\ast }(x)\varphi _{{\bf k%
}}(x)\varphi _{-{\bf k}}(x)$ are introduced in order to diagonalize the
resulting quadratic part $P_{O}^{-1}(k)O_{k}^{\ast
}O_{k}+P_{A}^{-1}(k)A_{k}^{\ast }A_{k}$, where $P_{O,A}^{-1}({\bf k,}%
k_{z})=2a+2bv^{2}(2\beta _{{\bf k}}\pm |\gamma _{{\bf k}}|)+k_{z}^{2}$ (to
simplify intermediate expressions an isotropic case $m_{ab}=m_{c}$ is
considered, results are generalized later). Functions $\gamma _{{\bf k}%
}=\lambda (-{\bf k},{\bf k})$ and $\beta _{{\bf k}}\equiv \int_{x}\varphi
^{\ast }(x)\varphi (x)\varphi _{{\bf k}}^{\ast }(x)\varphi _{{\bf k}%
}(x)=\lambda (0,{\bf k})$ as well as all the three and four leg vertices can
be expressed via single function of two quasimomenta 
\begin{equation}
\lambda ({\bf k}_{1},{\bf k}_{2})=\sum_{l,m}(-)^{lm}\exp \left\{ i\frac{2\pi 
}{a_{\bigtriangleup }}\left[ lk_{1}^{y}+mk_{2}^{y}\right] -\frac{1}{2}\left[
(k_{2}^{x}-\frac{2\pi }{a_{\bigtriangleup }}l)^{2}+(k_{1}^{x}-\frac{2\pi }{%
a_{\bigtriangleup }}m)^{2}\right] \right\}   \label{lambda}
\end{equation}
For example the $A_{k1}A_{k2}A_{-k1-k2}$ vertex is:

\begin{equation}
ibv%
\mathop{\rm Re}%
\left[ \lambda ({\bf k}_{1},{\bf k}_{2})\right] =\frac{ibv}{2}\beta
_{02}^{A}(k_{1}^{x}k_{1}^{y}+k_{2}^{x}k_{2}^{y})+O(k^{4}),  \label{supersoft}
\end{equation}
where $\beta _{st}^{A}\equiv \left( \frac{2\pi }{a_{\bigtriangleup }}\right)
^{4}\sum_{l,m}l^{s}m^{t}(-)^{lm}\exp \left[ -\frac{(2\pi )^{2}}{%
2a_{\bigtriangleup }^{2}}(l^{2}+m)^{2}\right] $. If the fluctuations were
absent the expectation value $v_{0}^{2}=\frac{a}{\beta _{A}b}$ would
minimize $G_{0}=-av^{2}+\frac{b}{2}\beta _{A}v^{4}$ where $\beta _{A}\equiv
\beta _{00}^{A}=1.16$. The propagators entering Feynman diagrams therefore
are: 
\begin{equation}
P_{O,A}(k)=\frac{1}{M_{O,A}^{2}({\bf k})+k_{z}^{2}};M_{O,A}^{2}({\bf k}%
)\equiv \frac{2a}{\beta _{A}}(-\beta _{A}+2\beta _{{\bf k}}\pm \gamma _{{\bf %
k}})  \nonumber
\end{equation}
Expanding around ${\bf k}=0$ using explicit expressions for $\gamma _{{\bf k}%
}$ and $\beta _{{\bf k}}$ one observes that constant and the $k^{2}$ terms
vanish, so that the (only) leading quartic term is $M_{A}^{2}({\bf k})=\frac{%
\beta _{22}^{A}}{2\beta ^{A}}|{\bf k}|^{4}$.

At one loop level the fluctuation contribution to the free energy is: 
\begin{equation}
G_{1}=\frac{1}{2}\frac{1}{(2\pi )^{3/2}}\int_{k_{z}}\int_{{\bf k}}\left\{
\log [P_{O}(k)]+\log [P_{A}(k)]\right\}   \label{corr}
\end{equation}
One should minimize $G_{0}+G_{1}$ with respect to $v$ leading to the
correction to its value: 
\begin{equation}
v_{1}^{2}=\frac{1}{(2\pi )^{3/2}}\int_{k_{z}}\int_{{\bf k}%
}[P_{O}(k)+P_{A}(k)]=\frac{1}{2(2\pi )^{1/2}}\int_{{\bf k}}\left[ \frac{1}{%
M_{O}({\bf k})}+\frac{1}{M_{A}({\bf k})}\right]   \label{corrvev}
\end{equation}

Due to additional softness of the $A$ mode the second ''bubble'' integral
diverges logarithmically in the infrared. This means that for the infinite
cutoff fluctuations destroy the inhomogeneous ground state, namely the state
with lowest energy is a homogeneous liquid \cite{gaugeinvphase}. Since the
divergence is logarithmic we are at lower critical dimensionality in which
an analog of Mermin - Wagner theorem \cite{Mermin} is applicable. It however
does not necessarily means that perturbation theory starting from ordered
ground state is inapplicable. The way to proceed in these situations have
been found while considering simpler models like $\varphi ^{4}$ model $F=%
\frac{1}{2}(\triangledown \varphi _{i})^{2}+V(\varphi _{i}^{2})$ in $D=2$
with number of components larger then 1, say $i=1,2$ \cite{Jevicki}.
Considering statistical sum, one first integrates exactly zero modes
existing due to continuous symmetry (translations in our case) and then
develops a perturbation theory via steepest descent method for the rest of
the variables. When the zero mode is taken out, there appears a single
configuration with lowest energy and steepest descent is well defined. For
invariant quantities like energy this procedure simplifies: one actually can
forget for a moment about integration over zero mode and proceed with the
calculation as if it is done in the ordered phase. The invariance of the
quantities ensures that the zero mode integration trivially factorizes. This
is no longer true for noninvariant quantities for which the machinery of
''collective coordinates method'' should be used \cite{Rajaraman}.

To the two loop order one gets several classes of diagrams, see Fig.1. The
leading order propagators are denoted by dashed and solid lines for the
''supersoft'' $A$ and ''hard'' $B$ modes respectively. The naively most
divergent diagram Fig.1D actually converges. To see this one writes
explicitly its expression in terms of function $\lambda $%
\begin{eqnarray}
&&\frac{b^{2}v^{2}}{2\left( 2\pi \right) ^{3/2}}\int_{q}\int_{p}I_{D}({\bf q}%
,{\bf p})P_{A}(p)P_{A}(q)P_{A}(p+q);  \label{diag} \\
I_{D}({\bf q},{\bf p}) &\equiv &-\lambda ({\bf p},-{\bf q})\lambda ({\bf p},%
{\bf q})+4\lambda ({\bf p}+{\bf q},{\bf p})\lambda ({\bf p}+{\bf q},{\bf q})%
\frac{\gamma _{{\bf p}+{\bf q}}}{\left| \gamma _{{\bf p}+{\bf q}}\right| }- 
\nonumber \\
&&-2\lambda ({\bf p}+{\bf q},-{\bf q})\lambda ({\bf p},-{\bf q})\frac{\gamma
_{{\bf p}}\gamma _{{\bf p}+{\bf q}}^{\ast }}{\left| \gamma _{{\bf p}}\gamma
_{{\bf p}+{\bf q}}\right| }++2\lambda ({\bf q},-{\bf p})^{2}\frac{\gamma _{%
{\bf p}}\gamma _{{\bf q}}\gamma _{{\bf p}+{\bf q}}^{\ast }}{\left| \gamma _{%
{\bf p}}\gamma _{{\bf q}}\gamma _{{\bf p}+{\bf q}}\right| }+c.c  \nonumber
\end{eqnarray}
The integrals over $\ p_{z}$ and $q_{z}$ can be explicitly performed using a
formula $\frac{1}{2\pi }\int_{p}\int_{q}\frac{1}{p^{2}+M_{1}^{2}}\frac{1}{%
q^{2}+M_{2}^{2}}\frac{1}{(p+q)^{2}+M_{3}^{2}}=\frac{\pi }{2}\frac{1}{%
M_{1}M_{2}M_{3}}\frac{1}{M_{1}+M_{2}+M_{3}}$. The leading divergence $\sim
\int_{{\bf p}}\int_{{\bf q}}I_{a}({\bf q},{\bf p})\frac{1}{{\bf p}^{2}{\bf q}%
^{2}\left| {\bf q}+{\bf p}\right| ^{2}}\frac{1}{{\bf p}^{2}+{\bf q}%
^{2}+\left| {\bf q}+{\bf p}\right| ^{2}}$, is determined by the asymptotics
of $I_{D}({\bf q},{\bf p})$ as both ${\bf p}$ and ${\bf q}$ approach zero.
If $I_{D}\sim 1$, it would diverge as $L^{4}$. However the vertex is
''supersoft'' at small quasimomenta $\sim p^{2}$ according to eq.(\ref
{supersoft}), so that expansion of $I_{D}({\bf q},{\bf p})$ starts from
terms quartic in ${\bf p}$ and ${\bf q}$ and there is no singularity at the
origin. This goes beyond the usual ''softness'' of interactions of the
Goldstone modes ($\sim p$). Nonleading divergences can be found by analyzing
contributions coming from three regions on which one of the line momenta $%
{\bf p}$, ${\bf q}$ or ${\bf p}+{\bf q}$ vanishes. The corresponding
expressions are $\sim \int_{{\bf k}}\int_{{\bf l}}I_{D}^{i}({\bf l})\frac{1}{%
{\bf k}^{2}M_{A}({\bf l})^{3}},$ with $I_{D}^{1}=0,I_{D}^{2}=\beta _{{\bf l}%
}^{2}-\beta _{{\bf l}}|\gamma _{{\bf l}}|$ and $I_{D}^{3}=-\beta _{{\bf l}%
}^{2}+\beta _{{\bf l}}|\gamma _{{\bf l}}|$ respectively. Here ${\bf k}$
denotes an IR divergent momentum while integration over ${\bf l}$ is
nonsingular. Although the second and the third contributions are divergent
their sum is convergent.

Standard methods similar to one used above can be applied to evaluate IR
divergencies of other superficially less divergent diagrams. There are no
power divergencies - only $\log ^{2}L$ and $\log L$.The results are $\frac{b%
}{\sqrt{2}\pi }\int_{{\bf p}}\frac{1}{M_{A}({\bf p})}\int_{{\bf q}}\left[ 
\frac{\beta _{A}-|\gamma _{{\bf q}}|}{M_{A}({\bf q})}+\frac{|\gamma _{{\bf q}%
}|}{M_{O}({\bf q})}\right] $, $\frac{b}{\sqrt{2}2\pi }\int_{{\bf p}}\frac{1}{%
M_{A}({\bf p})}\int_{{\bf q}}\frac{2\beta _{{\bf q}}+|\gamma _{{\bf q}%
}|-3\beta _{A}}{M_{A}({\bf q})}$, $\frac{b}{\sqrt{2}2\pi }\int_{{\bf p}}%
\frac{1}{M_{A}({\bf p})}\int_{{\bf q}}\frac{2\beta _{{\bf k}}-|\gamma _{{\bf %
q}}|}{M_{O}({\bf q})}$ for diagrams Fig1 a, b and e respectively. In
addition to direct contributions from $G_{2}$ Fig.1 there is also a
''correction term'' due to correction in the value of $v$ from eq.(\ref
{corrvev}) inserted into the lower order contributions to free energy $G_{0}$
and $G_{1}$. It's divergent part is $-\frac{b}{\sqrt{2}2\pi }\int_{{\bf p}}%
\frac{1}{M_{A}({\bf p})}\int_{\overrightarrow{q}}\left[ \frac{2\beta _{{\bf k%
}}-|\gamma _{{\bf q}}|-\beta _{A}}{M_{A}({\bf q})}-\frac{2\beta _{{\bf k}%
}+|\gamma _{{\bf q}}|}{M_{O}({\bf q})}\right] $. Both the leading
divergencies log$^{2}L$ and the next to leading ones log$L$ cancel between
the four contributions. Similar cancellations of all the logarithmic IR
divergencies occur in scalar models with Goldstone bosons in $D=2$ and $D=3$
(where the divergencies are known as ''spurious'').

The finite result for the Gibbs free energy to two loops (finite parts of
the integrals were calculated numerically) is restoring the original units
and reintroducing the asymmetry $m_{c}\neq m_{ab}$: 
\begin{equation}
G=\frac{\pi \hbar ^{2}}{eHk_{B}T\sqrt{m_{ab}}}g;g=-\frac{1}{2\beta _{A}}%
a_{T}^{2}+c_{1}\sqrt{|a_{T}|}+c_{2}\frac{1}{|a_{T}|}  \label{g}
\end{equation}
where numerical values of the coefficients are $c_{1}=3.16$ and $c_{2}=7.5$.
Dimensionless entropy (LLL scaled magnetization) 
\begin{equation}
s=-\frac{dg}{da_{T}}=\left( \frac{\pi ^{2}c^{5}m_{ab}^{3}b}{%
8e^{5}k_{B}^{2}m_{c}}\right) ^{1/3}\frac{M}{(TH)^{2/3}}=\frac{1}{\beta _{A}}%
a_{T}+\frac{c_{1}}{2}\frac{1}{|a_{T}|}-c_{2}\frac{1}{a_{T}^{2}}  \label{s}
\end{equation}
and specific heat normalized to the mean field value 
\begin{equation}
\frac{1}{\beta _{A}}\frac{C}{\Delta C}=-\frac{d^{2}g}{da_{T}^{2}}=1+\frac{%
c_{1}}{4}\frac{1}{|a_{T}|^{3/2}}+2c_{2}\frac{1}{a_{T}^{3}}  \label{c}
\end{equation}
for successive partial sums are plotted on Fig.2 and 3 (dashed lines).
Qualitative they are in accord with numerous experiments and MC simulations 
\cite{Sasik}. At low temperature magnetization is a bit larger then the mean
field's one, while dimensionless specific heat characteristically grows
before dropping fast around $a_{T}=-5$. To make more detailed comparison, I
interpolated between results of low temperature expansion and these of high
temperature expansion using the following rational form for free energy in
terms of often used variable $x$ defined implicitly by $%
x=y^{2},a_{T}=4(2y)^{2/3}\left( 1-1/8y^{2}\right) $: 
\begin{equation}
g=4(2y)^{2/3}\frac{1+a_{1}y+...+a_{n+2}y^{n+2}}{1+b_{1}y+...+b_{n}y^{n}}
\label{interpol}
\end{equation}
The coefficients were constrained from both low and high temperature sides.
It has been already noted \cite{Moore1} that constraining from both sides
the Pade approximants just by the first term\bigskip\ at low energy improves
otherwise unsatisfactory magnetization and specific heat. Adding two more
terms on the low temperature end makes it very close to the MC results
(stars,triangles and diamonds correspond to 1T, 2T and 5T results for YBCO).
I used just three leading terms in high temperature expansion shown on Fig.2
and 3 by dotted lines. Using more terms does not modify significantly the
result. Although magnetization curve eq.(\ref{interpol}) agrees with that of
ref.\cite{Tesanovic}, the specific heat is not.

To summarize, it is established up to order of two loops that perturbation
theory around Abrikosov lattice is consistent. All the IR divergencies
cancel due to soft interactions of the soft mode. Perturbative results as
well as interpolation with the high temperature expansion agree very well
with the direct MC simulation.

Now I comment on range of validity of the perturbative results and
nonperturbative effects. As can be seen from Fig.2 and 3 the range of
validity of the low temperature expansion presented in this paper is below $%
a_{T}=-10$, while that of the high temperature expansion is above $a_{T}=-2$%
. Both exclude the range in which small magnetization jump (not seen on
Fig.2's scale) due to vortex melting is seen experimentally and in the
numerical simulation. Since the MC simulation is the only systematic tool
available in the intermediate region (the theory of Te\v{s}anovi\'{c} {\it %
et al} \cite{Tesanovic} captures major (98\%) contribution, but does dot
treat the small (2\%) effect including melting), one might have two
possibilities to discuss such a singularity within the present framework.
One possibility is that the jump is due to finite size effects and
disappears in the thermodynamic limit (value of the cutoff in the simulation
is only $L\sim 25$). Another is that some nonperturbative effects can
stabilize the vortex lattice. Quantitative comparison with experiments on
YBCO was attempted in ref. \cite{Sasik,Tesanovic}. The present simple
interpolation formula eq.(\ref{interpol}) works equally well.

Author is very grateful to L. Bulaevsky for encouragement and numerous
discussions, R. Sasik for providing original MC data and explaining details
of his MC simulation, Y. Kluger and other members of T11 and T8, especially
A. Balatsky for discussions and hospitality in Los Alamos where part of this
work was done. Work was supported by grant NSC of Taiwan.

\newpage

{\large Figure Captions}

Fig.1.

Contributions to the free energy at the two loop level.

Fig. 2.

Scaled magnetization defined in eq.(\ref{s}). Dashed (dotted) lines are
successive low (high) temperature approximants, while the solid line is the
interpolation. Points are the MC results.

Fig. 3

Scaled specific heat defined in eq.(\ref{c}). Same notations as in Fig.2.


\begin{references}
\bibitem{gaugeinvphase}  Note that this (gauge invariant) argument is
completely independent of choice of the ''order parameter'' or ''gauge
invariant phase'' see [7] and references therein.
\end{references}
\end{document}